\documentclass[structabstract]{aa}
\usepackage[pdftex,final]{graphicx}
\usepackage{txfonts}
\usepackage{natbib}
\usepackage{multirow}
\usepackage{subfig}
\bibpunct{(}{)}{;}{a}{}{,} 

\begin{document}
%
   \title{COSMOGRAIL: the COSmological MOnitoring of GRAvItational Lenses \thanks{Based on observations made with the 1.2-m Swiss Euler telescope (La Silla, Chile), the 1.5-m AZT-22 telescope (Maidanak Observatory, Uzbekistan), the 2.0-m HCT telescope (Hanle, India), and the 1.2-m Mercator Telescope. Mercator is operated on the island of La Palma by the Flemish Community, at the Spanish Observatorio del Roque de los Muchachos of the Instituto de Astrof\'isica de Canarias.}\fnmsep\thanks{Light curves \emph{will be} available at the CDS via anonymous ftp to cdsarc.u-strasbg.fr (130.79.128.5) or via http://cdsarc.u-strasbg.fr/viz-bin/qcat?J/A+A/???, and on http://www.cosmograil.org.}}

   \subtitle{XII. Time delays of the doubly lensed quasars SDSS~J1206+4332 and HS~2209+1914 }

\author{E.Eulaers
\inst{1}
\and M. Tewes
\inst{2}
\and P. Magain 
\inst{1}
\and F. Courbin 
\inst{2}
\and I. Asfandiyarov
\inst{3}
\and Sh. Ehgamberdiev
\inst{3}
\and S. Rathna Kumar
\inst{5}
\and C.S. Stalin
\inst{5}
\and T.P. Prabhu
\inst{5}
\and G. Meylan
\inst{2}
\and H. Van Winckel
\inst{4} 
}

\institute{Institut d'Astrophysique et de G\' eophysique, Universit\' e de Li\`ege, All\'ee du 6 Ao\^ut, 17, Sart Tilman (Bat. B5C), Li\`ege 1, Belgium
\email{E.Eulaers@ulg.ac.be}
\and
Laboratoire d'Astrophysique, Ecole Polytechnique F\' ed\' erale de Lausanne (EPFL), Observatoire de Sauverny, CH-1290 Versoix, Switzerland
\and
Ulugh Beg Astronomical Institute, Uzbek Academy of Sciences, Astronomicheskaya 33, Tashkent, 100052, Uzbekistan
\and
Instituut voor Sterrenkunde, Katholieke Universiteit Leuven, Celestijnenlaan 200B, 3001 Heverlee, Belgium
\and
Indian Institute of Astrophysics, II Block, Koramangala, Bangalore 560 034, India
}

\date{}

 
  \abstract
   {}
   {Within the framework of the COSMOGRAIL collaboration we present 7- and 8.5-year-long light curves and time-delay estimates for two gravitationally lensed quasars: SDSS~J1206+4332 and HS~2209+1914.}
   {We monitored these doubly lensed quasars in the R-band using four telescopes: the Mercator, Maidanak, Himalayan Chandra, and Euler Telescopes, together spanning a period of 7 to 8.5 observing seasons from mid-2004 to mid-2011. The photometry of the quasar images was obtained through simultaneous deconvolution of these data. The time delays were determined from these resulting light curves using four very different techniques: a dispersion method, a spline fit, a regression difference technique, and a numerical model fit. This minimizes the bias that might be introduced by the use of a single method.}
   {The time delay for SDSS~J1206+4332 is $\Delta t_{AB} = 111.3 \pm 3$ days with A leading B, confirming a previously published result within the error bars. For HS~2209+1914 we present a new time delay of $\Delta t_{BA} = 20.0 \pm 5$ days with B leading A.}
   {The combination of data from up to four telescopes have led to well-sampled and nearly 9-season-long light curves, which were necessary to obtain these results, especially for the compact doubly lensed quasar HS~2209+1914.}

   \keywords{Gravitational lensing: strong --
	  Methods: numerical --
          Galaxies: quasars: individual: SDSS~J1206+4332, HS~2209+1914 --
	  Cosmology: cosmological parameters --
               }
   \titlerunning{XIII. Time delays of SDSS~J1206+4332 and HS~2209+1914}
   \maketitle
 
%

\section{Introduction}

\citet{Refsdal1964MNRAS_H0} was the first to realize that the determination of the time delay between different images of a gravitationally lensed source can be used to derive the Hubble constant $H_{0}$, the expansion rate of the Universe, if the mass distribution in the lensing galaxy is known. The other way around, if a value for $H_{0}$ is assumed, the time delay between the lensed quasar images allows one to study the mass of the lensing galaxy. 

Even if this method presents the advantage of being independent of other calibrations, it has often been subject to controversies both on time-delay values \citep[e.g. on QSO 0957+561 see ][]{Press1992a,Press1992b,Pelt1996AA,Kundic1997} and on the Hubble constant $H_{0}$ derived from these time delays, this value seeming lower than the values derived by other techniques \citep{Burud2001PhDT}. A decade ago, time delays were measured on relatively short light curves of only a few seasons \citep[e.g. ][]{Burud2002AA_HE2149,Patnaik2001,Corbett1996IAUS_B0218}, using only one time-delay extracting method at a time. This could lead to biased results and an underestimation of the error bars as pointed out by \citet{Eulaers2011}. Nowadays, the time-delay approach is considered to be competitive and highly complementary to other cosmological probes, and it gives similar results for $H_{0}$ \citep{Oguri2007,Linder2011,Suyu2012a}. 

COSMOGRAIL is a project aimed at measuring time delays in most known gravitationally lensed quasars. The use of 1-m to 2-m class semi-dedicated telescopes has allowed us to acquire light curves spanning up to ten years of data. In combination with a homogeneous way of analysing data as described in \citet[][accepted]{Tewes2012TD} and \citet[][submitted]{Tewes2012_RXJ1131}, the collaboration is contributing to a significant rise in reliable time-delay measurements of gravitationally lensed quasars. The first results included time delays for SDSS~J1650+4251 \citep{Vuissoz2007}, and WFI~J2033-4723 \citep{Vuissoz2008} and more recently for HE~0435-1223 \citep{Courbin2011} and RX~J1133-1231 \citep[][ submitted]{Tewes2012_RXJ1131}. In this paper, we present time delays for two doubly lensed quasars: SDSS~J1206+4332 ($12^{h}06^{m}29^{s}.65, +43\degr32\arcmin17\farcs60; J2000.0$) and HS~2209+1914 ($22^{h}11^{m}30^{s}.30, +19\degr29\arcmin12\farcs00; J2000.0$).

SDSS~J1206+4332 was discovered by \citet{Oguri2005}. They found a $z=1.789$ quasar whose lensed images are separated by $2\farcs90$. The redshift of the main lensing galaxy is estimated at $z=0.748$. One or two other nearby galaxies might contribute to the lensing potential. 

During the Hamburg Quasar Survey \citet{Hagen1999} identified the system HS~2209+1914 as a bright $z_{s}=1.07$ quasar having a galaxy neighbour detectable on their Schmidt plates. Nothing else was published on this object until \citet{Chantry2010}, who obtained precise astrometry for the two lensed quasar images and the lensing galaxy based on the deconvolution of Hubble Space Telescope images.

In Section 2 we summarize the data acquisition and reduction techniques, and briefly explain the methods used to determine the time delays. In Section 3, we present the results for the doubly lensed quasar SDSS~J1206+4332, including photometry and time delays. In Section 4, we take a closer look at HS~2209+1914 for which we give a first time-delay estimate. The last section is a summary of our results. 


\section{Data acquisition and reduction}
\label{sec:photo}

\subsection{Data acquisition}

Both lensed quasars were monitored for more than seven years on different telescopes.
The 1.2-m Mercator telescope, located at the Roque de los Muchachos Observatory on La Palma, Canary Islands (Spain), is equipped with the MEROPE camera having a field of view of $6\farcm5$ by $6\farcm5$ and a scale of $0\farcs19$ per pixel. COSMOGRAIL observations by the Mercator telescope stopped in December 2008.
The remotely operated 2m Himalayan Chandra Telescope (HCT) started systematic monitoring for the COSMOGRAIL collaboration in 2007 and has continued up to now. It is equipped with the Hanle Faint Object Spectrograph Camera (HFOSC), a $2048 \times 4096$ pixels CCD camera with a field of view of $10\arcmin$ by $10\arcmin$ and a scale of $0\farcs296$ per pixel.

These two telescopes provided the bulk of the data, and were complemented by data from the 1.5-m telescope at Maidanak Observatory, Uzbekistan. Most of the data were obtained with the SITE CCD camera, a $800 \times 2000$ array camera with a scale of $0\farcs266$ per pixel and a field of view of $3\farcm5 \times 8\farcm9$. During two seasons, a second camera (SI) was used with an array of $4096 \times 4096$ pixels and a field of view of $18\farcm1$ by $18\farcm1$ on the same scale of $0\farcs266$ per pixel.

For the lensed system HS~2209+1914, some data were added from the 1.2-m Euler telescope, located at La Silla, Chile. It is equipped with a $2048 \times 2048$ CCD camera having a field of view of $11 \arcmin$ by $11\arcmin$ and a scale of $0\farcs344$ per pixel.

At every observing epoch, and as long as there were no technical or meteorological problems, five dithered images were taken in the $R$-band with an exposure time of six minutes each.


\subsection{Data reduction}

To analyse COSMOGRAIL data in a consistent and homogeneous way, we use the semi-automated pipeline that was described in \citet{Tewes2012_RXJ1131} (submitted) to carry out the reduction of the CCD images. We summarize the main points after the bias subtraction and flatfielding of the frames.


Four stars are used to construct a point spread function (PSF) per frame. Photometry of the sources is obtained through simultaneous deconvolution using the MCS algorithm \citep{Magain1998}, where the main feature is the finite final resolution, which avoids artefacts. The deconvolved images have a pixel scale of half the telescope's pixel scale and have a Gaussian PSF with a two pixel full-width-at-half-maximum. The flux of the point sources is allowed to vary from one frame to another, unlike the lensing galaxy, which is part of the numerical background that is held fixed throughout the frames. The simultaneous deconvolution of all frames constrains the positions of the quasar images very well, even if not all data have been obtained under optimal seeing conditions. 


An estimation of the photometric shot noise $\sigma_{N}$, calculated for each quasar image and frame, is given by \footnote{Based upon Heyer, Biretta, 2005, WFPC2 Instrument Handbook, Version 9.1, Chapter 6}

\begin{eqnarray}
\sigma_{N} = \sqrt{f_{*} + \frac{N_{sky} + R^{2}}{S}}
\end{eqnarray}
with $f_{*}$ the flux in the quasar image (expressed in number of photons), $N_{sky}$ the sky level of the exposure, $R$ the CCD read noise (both in photons per pixel), and $S$ the PSF sharpness given by
\begin{eqnarray}
S = \sum\limits_{ij} (PSF_{ij})^{2},
\end{eqnarray}
i.e. the sum over all pixels of the squared normalized fluxes of the PSF pixels.


The data points in the light curves are the median photometry per night obtained by the simultaneous deconvolution of the images. Their error bars are the maximum of two kinds of error bars: an empirical error bar and an intrinsic one. Indeed, the total error on the night cannot be smaller than the one given by the spread of the measurements in that night, nor can it be smaller than the error bar given by the photon noise combined with the renormalization error, so taking the maximum of both is justified. They are generally very close to one another.

The empirical error reflecting the spread of the measurements during one night is estimated by the standard deviation of the measurements per night normalized by the number of images per night. Since our photometry is a median value per night, the standard deviation is estimated via the median absolute deviation (MAD) estimator \citep{Hoaglin} corrected by a scale factor for normal distributions. The MAD is defined as the median of the absolute values of the residuals from the data's median:
\begin{eqnarray}
MAD = med_{i} \left( \vert X_{i} - med_{j}(X_{j}) \vert \right) 
\end{eqnarray}
It is related to the standard deviation via $\sigma = K * MAD$.  In case of a normal distribution, $K = \frac{1}{\phi^{-1}(0.75)} \approx 1.4862$, with $\phi^{-1}$ being the quantile function of the normal distribution.
 
On the other hand, the intrinsic or theoretical error, calculated as the quadratic sum of the median shotnoise per night and the median renormalization error per night, gives a theoretical estimate of the error on the combined data point. That neither the empirical nor the theoretical error systematically dominate is a proof of the coherence of both approaches, and it justifies the idea of taking the maximum of both. 

Finally, we put together the light curves of the same object coming from different telescopes by optimizing the magnitude shifts between the curves of the same image from different instruments.

\subsection{Time-delay analysis}

To avoid a bias due to the method, we use four different techniques for our time-delay analysis. Three of them, a spline fit, a dispersion minimization, and a method based on the variability of regression differences, are described and tested extensively in \citet{Tewes2012TD} (submitted). The fourth method, the numerical model fit, was described and applied in \citet{Eulaers2011}. We quickly summarize the main features of each of them:

\begin{enumerate}
\item {\bf The dispersion-like technique} considers the time delay to be the shift between the light curves that minimizes the dispersion between these data. Microlensing is modelled by polynomials up to the third order, but no model light curve for the overall quasar variability is constructed here. 
\item {\bf The regression difference technique} searches for the time delay that minimizes the variability of the differences between the time-shifted continuous regressions of the light curves. Microlensing variability is not explicitly modelled here. 
\item {\bf The free knot spline technique} models the intrinsic quasar variability, as well as the microlensing variations, by fitting splines simultaneously to the light curves. The curves are shifted in time to optimize this fit.
\item{\bf The numerical model fit (NMF)} constructs a numerical model for the quasar variability, together with a linear microlensing trend for a given time delay. The optimal time delay is the one that minimizes the difference between the data and this numerical model.
\end{enumerate}

These four techniques are based on very different principles, so we expect them to be sensitive to different sources of error. By applying all of them to our light curves, we minimize the bias that might be due to the choice of the method.

We use a Monte Carlo approach for the time-delay uncertainty calculations. For the first three methods, these are based on simulated light curves with known time delays based on the spline fit as explained in \citet{Tewes2012TD} (submitted). For the last method, we use the numerical model light curve to which we add appropriate Gaussian noise.

\section{SDSS~J1206+4332}

\subsection{Deconvolution results and light curves}

For SDSS J1206+4332 we use data coming from three telescopes. At Maidanak this lensed quasar was observed from January 2005 until July 2008. During the 2007 season, the second camera SI was used. Mercator data span the period March 2005 till December 2008, and HCT has continued monitoring this lens since May 2007 until now. 
The combination of these data results in seven complete observing seasons from 2005 until 2011. 

The object is unvisible from August until the beginning of November, a period of more or less 100 days. The mean temporal sampling during the visibility window is about one point every fourth night for Mercator and Maidanak data, and one point per week for HCT. In the overlapping 2008 season this gives more than one point every second night.
Table \ref{tab:J1206_monitoring} gives an overview of the optical monitoring for this quasar. One epoch corresponds to one data point in the light curve, which is the median value for that night as explained in Section \ref{sec:photo}. S is the temporal sampling of the observations expressed as an approximative average in number of days excluding the seasonal gaps.

\begin{table*}[htdp]
\caption{Overview of the optical monitoring for SDSS~J1206+4332. 
}
\begin{center}
\begin{tabular}{lllcclrr}
\hline
\hline
Telescope & Location & Instrument & Pixel scale & Field & Time span  & Epochs & S \\
\hline
Mercator 1.2-m & La Palma, Canary Islands, Spain  & MEROPE & $0\farcs193$ & $6\farcm5 \times 6\farcm5$ & Mar 2005 - Dec 2008 & 196 & $\backsim 4d$\\
AZT-22 1.5-m & Maidanak, Uzbekistan  & SITE & $0\farcs266$ & $3\farcm5 \times 8\farcm9$ & Jan 2005 - July 2008 & 57 & $\backsim 4d$ \\
AZT-22 1.5-m & Maidanak, Uzbekistan  & SI & $0\farcs266$ & $18\farcm1 \times 18\farcm1$ & 2007 & 14 & $\backsim 4d$\\
HCT 2m & Hanle, India & HFOSC &  $0\farcs296$ & $10\arcmin \times 10\arcmin$ & May 2007 - July 2011 & 115 & $\backsim 7d$\\
\hline
\end{tabular}
\end{center}
\label{tab:J1206_monitoring}
\end{table*}

\begin{figure*}[htbp]
\begin{center}
\includegraphics[width=0.95\linewidth]{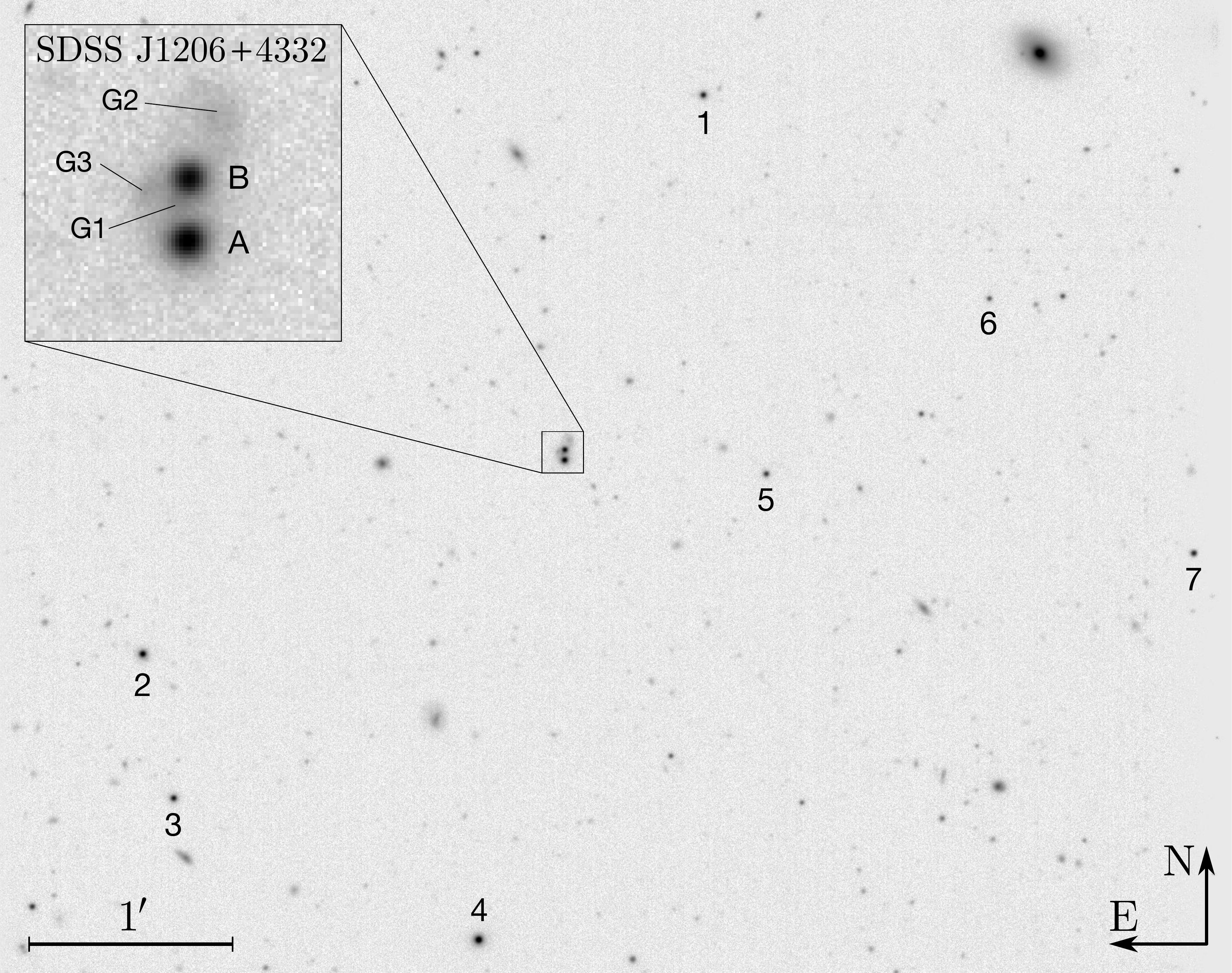}
\caption{Part of the field around SDSS~J1206+4332. It is a combination of 29 frames taken at the 1.2-m Mercator telescope with a seeing $\leq 1\farcs2$ and an ellipticity $\leq 0.12$. The stars used to model the PSF are labelled 1 to 4. In case one of these stars did not have sufficient quality, three other stars labelled 5 to 7 were used. North is up and east is left.
}
\label{fig:J1206_ref}
\end{center}
\end{figure*}

The four stars, designated 1 to 4 in Figure \ref{fig:J1206_ref}, are used to model the PSF needed for the simultaneous deconvolution. If these stars did not satisfy our quality criteria (due to e.g. cosmic ray hits), a second set of four stars was used: star 1 in combination with star 5 to star 7. Five stars, 1 to 5, were also used as flux calibrators.

\begin{figure}[htbp]
\begin{center}
\includegraphics[width=0.9\linewidth]{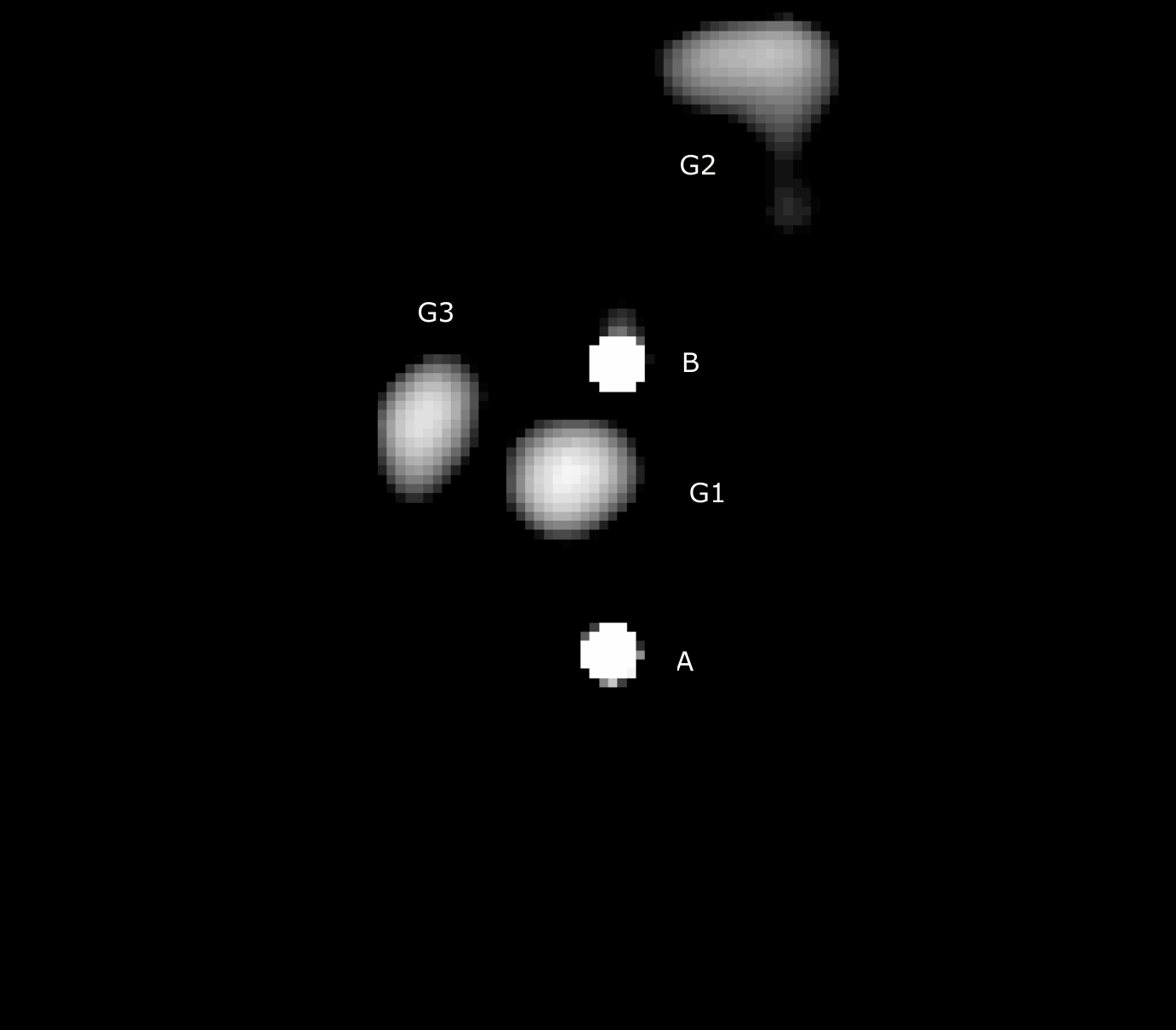}
\caption{Result of the simultaneous deconvolution of 1109 R-band Mercator images of J1206+4332. The pixel size is $0\farcs095$ (i.e. half the pixel size of the Merope CCD camera at the Mercator telescope), and the resolution is 2 pixels full-width-half-maximum. In addition to the main lensing galaxy $G1$ between the lensed images $A$ and $B$ we clearly identify two more galaxies: $G2$ to the north of the system and $G3$ to the east. North is up and east is left.}
\label{fig:J1206_dec}
\end{center}
\end{figure}

Figure \ref{fig:J1206_dec} shows that the simultaneous deconvolution of 1109 R-band Mercator images clearly reveals the lens galaxy between the lensed quasar images, as well as two other galaxies close to the system, which have also been identified by \citet{Oguri2005}.  
Table \ref{tab:J1206_astrometry} presents the relative astrometry of the system as determined from this deconvolution. It shows a slightly different position for the B image and the galaxy G1 in comparison with the positions in \citet{Oguri2005}.

\begin{table}[htdp]
\begin{center}
\begin{tabular}{lrr}
\hline
\hline
Object & $x$ (arcsec) & $y$ (arcsec) \\
\hline
A & $0.000 \pm 0.004$ & $0.000 \pm 0.004$\\
B & $ 0.073 \pm 0.005$ & $2.970 \pm 0.004$\\
G1 & $-0.411 \pm 0.042$ & $1.817 \pm 0.012$ \\
G2 & $1.482 \pm 0.006$ & $5.959 \pm 0.010$ \\
G3 & $-1.902 \pm 0.027$ & $2.353 \pm 0.021$ \\
\hline
\end{tabular}
\end{center}
\caption{Astrometry for SDSS~J1206+4332 relative to the brighter quasar image A. The positive directions for $x$ and $y$ are west and north respectively.}
\label{tab:J1206_astrometry}
\end{table}

We obtain an image separation of $2\farcs98$ between the quasar images A and B, which is slightly higher than \citet{Oguri2005}, but a lower value for the flux ratio $F_{A}/F_{B} =1.31$, for which \citet{Oguri2005} found $1.48$. Our value is the result of the deconvolution and has not been corrected for the time delay yet. We come back to this in the next section.

\begin{figure*}[htbp]
\resizebox{\hsize}{!}{\includegraphics{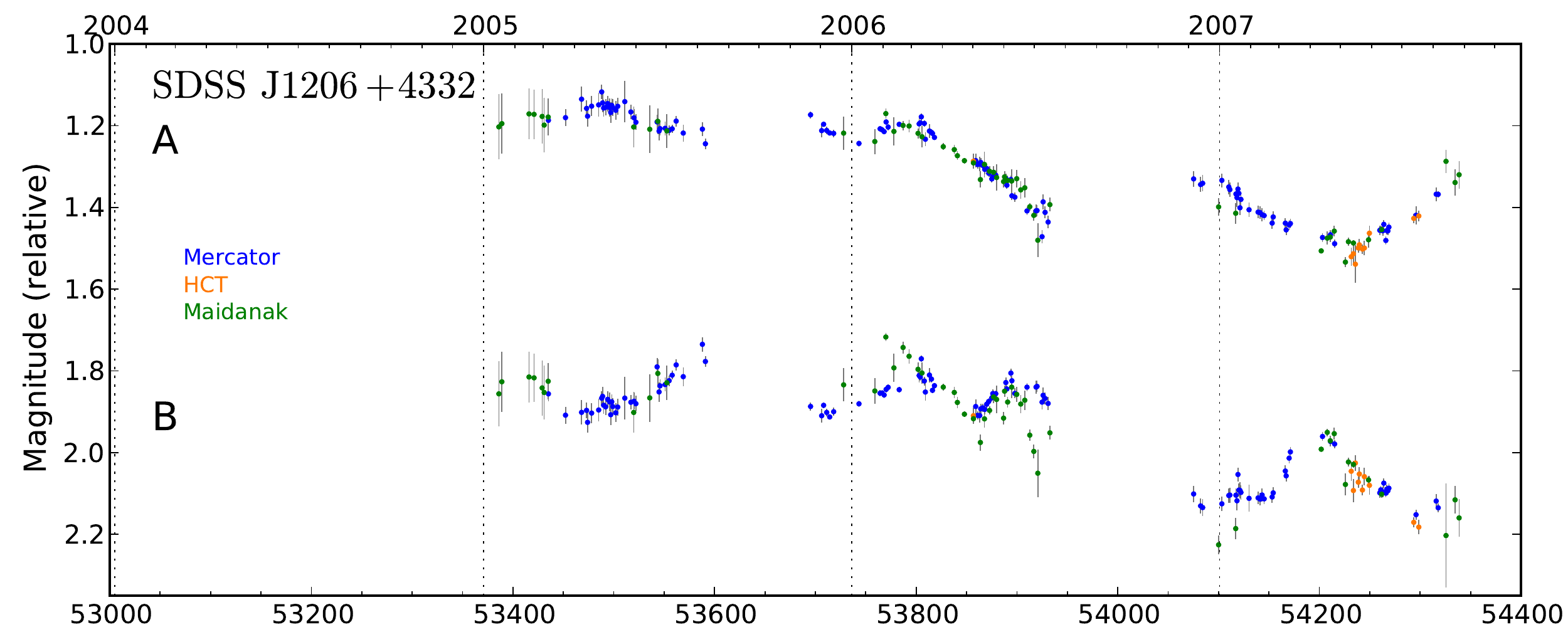}}
\resizebox{\hsize}{!}{\includegraphics{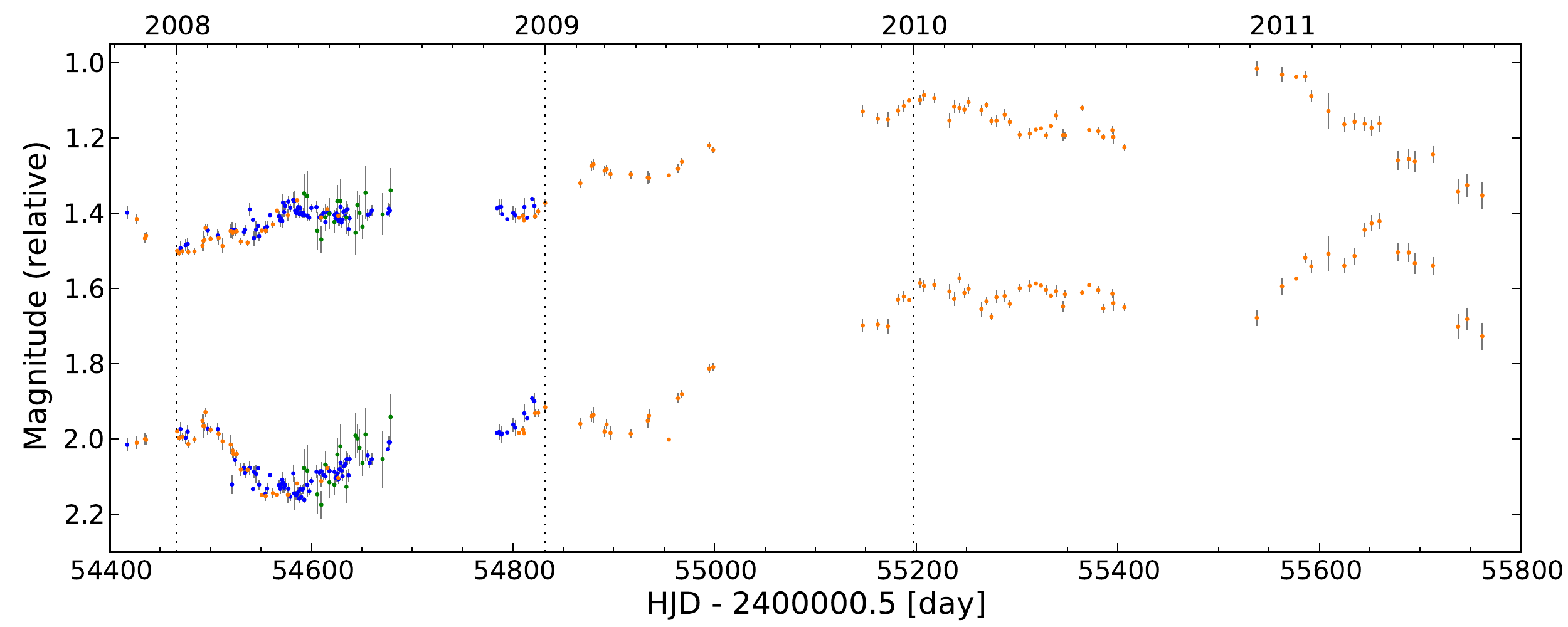}}
\caption{Light curves of SDSS~J1206+4332 combining data from Mercator, Maidanak, and HCT, and spanning 7 observing seasons. Photometry is obtained from simultaneous deconvolution, and the $1\sigma$ error bars are calculated as described in Section \ref{sec:photo}.}
\label{fig_lcJ1206}
\end{figure*}

The light curves that have been obtained as described in Section \ref{sec:photo} are shown in Figure \ref{fig_lcJ1206}. The 2007 and 2008 seasons include data from all three telescopes and show that they match and complement each other well. We distinguish clear features of quasar variability over periods as short as two months, for example in the 2006 season of the B curve. Visually, the light curves are compatible with a time delay around $\backsim 100$ days. Finally, we also observe an evolution of the flux ratio between the quasar images, especially in the 2010 and 2011 seasons.

\subsection{Time delay}
\label{subsec:td_J1206}

Based on their lens modelling, \citet{Oguri2005} predicted a time delay for SDSS~J1206+4332 of $\Delta t_{AB} = 92.6$ or $104.4$ days depending on the influence of the secondary galaxy G2. With only one observing season, \citet{Paraficz2009} found a time delay of $\Delta t_{AB} = 116_{-5}^{+4}$ days.

\begin{figure}[htbp]
\begin{center}
\includegraphics[width=0.99\linewidth]{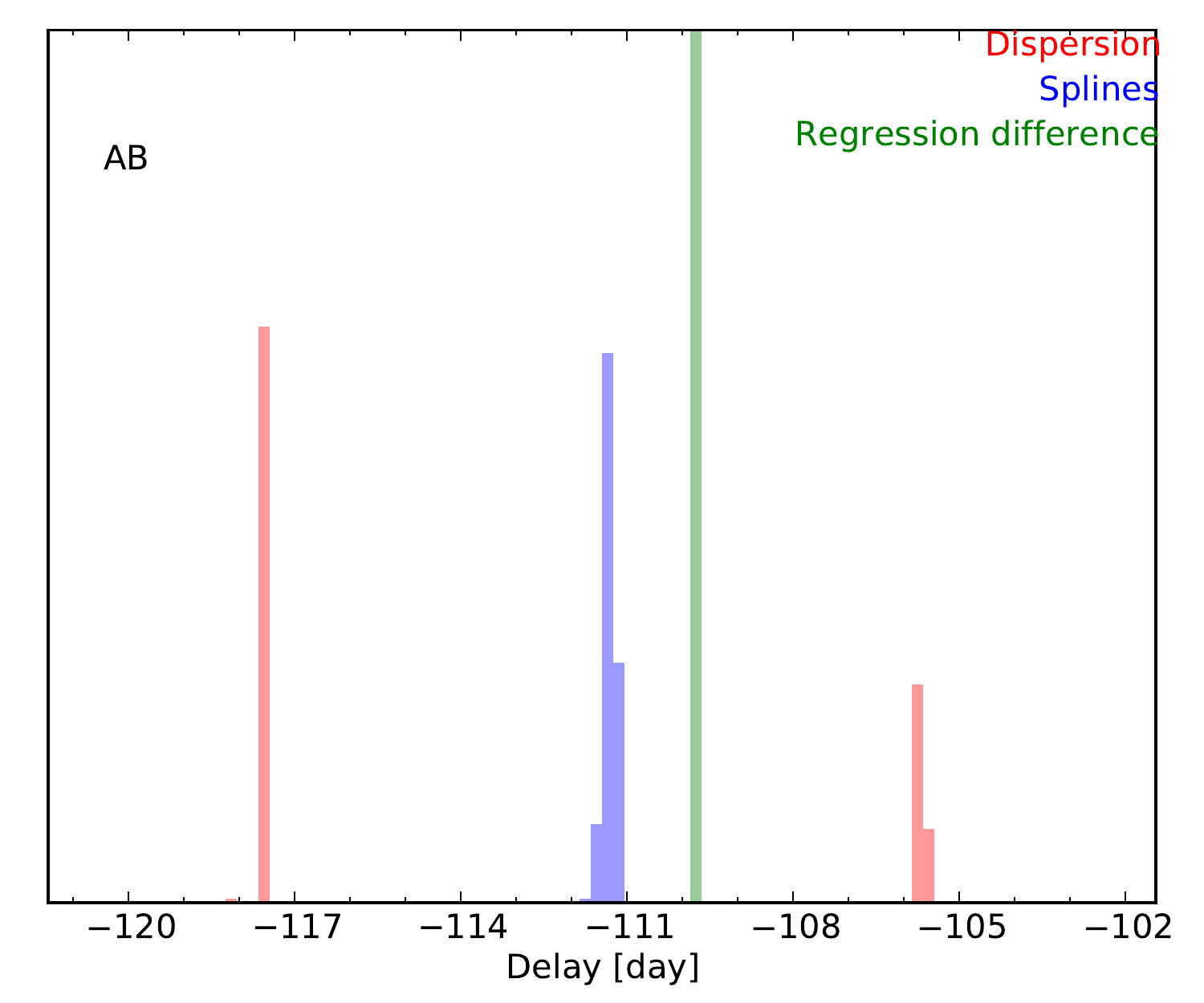}
\caption{Distribution of time delays for SDSS~J1206+4332 
by randomly varying 200 times the initial time delay needed for three of our methods. Only the dispersion technique shows a secondary peak.}
\label{fig:J1206_intrinsvar}
\end{center}
\end{figure}

Our time-delay analysis is based on the light curves presented in Figure \ref{fig_lcJ1206}, and it uses the four different methods briefly explained in Section \ref{sec:photo}.
Since three of these techniques (dispersion, regression difference and spline fit) rely on optimizations of a first guess of the time delay, we test the robustness of the result by running 200 times these techniques on the same data set, while uniformly and randomly varying the initial delay in a range of $\pm 10$ days around the first guess. Figure \ref{fig:J1206_intrinsvar} shows that in the case of SDSS~J1206+4332 we have monomodal distributions for the spline and regression techniques, and a bimodal distribution for the dispersion technique, which is inherently more sensitive to the variation of this entry delay. 
The results of this test confirm that our method parameters are well adapted to the data, 
so that we can use the mean value of these distributions as our final best time delay.

\begin{figure}[htbp]
\begin{center}
\includegraphics[width=0.9\linewidth]{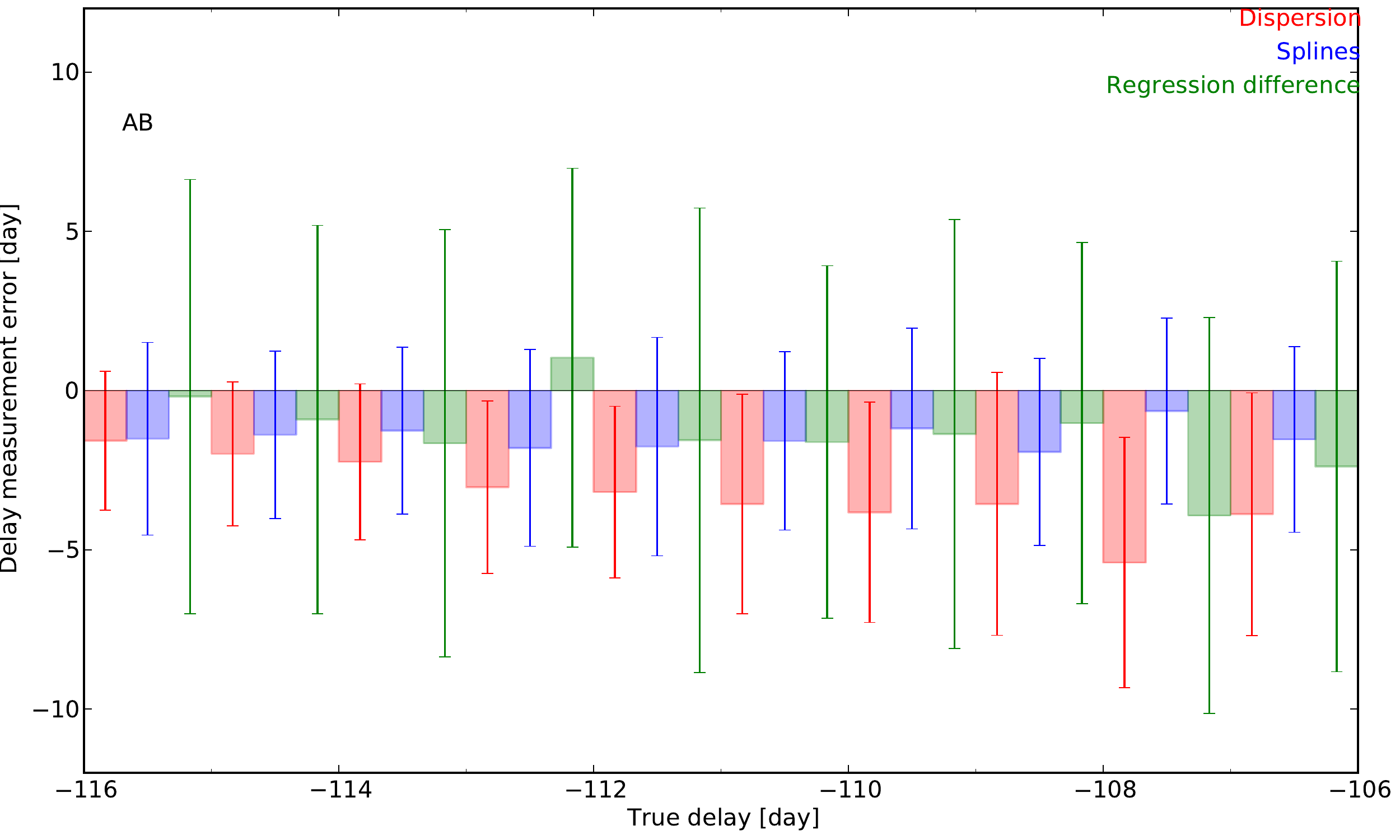}
\caption{Error analysis of the simulated light curves for SDSS~J1206+4332 using data from Mercator, HCT, and Maidanak. We plot binned mean measurement errors and their standard deviations against the true delays of the synthetic curves. The systematic error is largest for the dispersion technique, and random errors dominate for the regression and spline methods. }
\label{fig:J1206_measvstrue}
\end{center}
\end{figure}

To obtain realistic error bars, we applied these three methods to 1000 synthetic light curves with known time delays. The error bars for these three techniques consist of a systematic and a random part, which then are summed quadratically to obtain the total error. Figure \ref{fig:J1206_measvstrue} illustrates both contributions: we plot the mean measurement error as a function of the true delay. The error bars are the standard deviation of these measurement errors. To be conservative, the maximum measurement error is our systematic error, and the maximum standard deviation our random error, which are then summed quadratically. 

The numerical model fit (NMF) is completely independent of the three other methods in how it determines the time delay and its associated error bars. We add normally distributed random errors with the appropriate standard deviation to the numerical model light curve and redetermine the time delay. This procedure is repeated 1000 times. The mean value of the time-delay distribution that we obtain in that way is considered to be the final time delay, and its dispersion represents the 1$\sigma$ error bar.  Because the errors added to the model, hence the final error bars on the time delay, depend on the photometric error bars of our data, we checked that our residuals (data minus model) were compatible with a normalized Gaussian distribution.

One of the parameters to be optimized in the NMF method is the flux ratio corrected by the time delay, for which we find $F_{A}/F_{B} = 1.45$ for the complete data set, which is very close to the \citet{Oguri2005} value of $1.48$.

We present an overview of the time delays, along with 1-$\sigma$ error bars, obtained by the different methods in Table \ref{tab:TD_J1206}, first for the Mercator and HCT data, two sets that really complement each other, and then for the whole data set coming from the three telescopes. Adding Maidanak data to the Mercator and HCT light curve adds more scatter, but seems to constrain the model light curves better, which is reflected in larger error bars for the dispersion method that does not involve a model, but smaller ones for the regression difference technique, the spline fit, and the numerical model fit. Image A leads image B by $\Delta t_{AB}$ days. 
Our results coming from four basically very different time-delay estimation techniques are very consistent, and confirm the existing estimation from \citet{Paraficz2009}, while improving its error bars for the spline fit and the NMF techniques used on the complete data set.

\begin{table*}[htbp]
\begin{center}
\begin{tabular}{lll}
\hline
\hline
Method & Mercator +HCT & Mercator + HCT + Maidanak \\ 
\hline

Dispersion & $111.52 \pm 4.95$ (4.57, 1.89) & $113.65 \pm 6.79$ (4.12, 5.40) \\
Regression  & $110.89 \pm10.04$ (9.55, 3.08) & $109.73 \pm 8.28$ (7.29, 3.92) \\
Spline & $111.18 \pm 5.79$ (5.00, 2.92) & $111.31 \pm 3.93$ (3.43, 1.92) \\
NMF & $111.87 \pm 0.96$ & $113.80 \pm 0.90 $\\

\hline
\end{tabular}
\end{center}
\caption{Summary of time delays for SDSS~J1206+4332 obtained by the four different methods. The indicated error is the total error. We mention between brackets the random and the systematic error for the first three methods as explained in the text.}
\label{tab:TD_J1206}
\end{table*}

That the error bars for the dispersion technique, the regression difference, and the spline fit are systematically larger than for the NMF is inherent to the way they are calculated. For these first three methods, the error bars are the quadratic sum of two maximum contributions, as described before, whereas the NMF does not take the systematic error into account (generally under one day), and considers the standard deviation of the time delay to be its random error without taking a maximum over models with slightly different time delays. On top of that, the random error contribution of the first three methods is the result of red noise, whereas the NMF only takes white noise into account.

Since the time delays resulting from the four methods all converge towards the same value, we only have to decide on the size of the error bars, which reflect the different sensitivities to various aspects that are present in the data and the method. 
Given the low spread in time-delay values between the methods, and given the possibility that the error bars from the dispersion methods, the regression difference technique, and the spline fit have been overestimated, while those for the NMF might be underestimated, we think that a final time delay for SDSS~J1206+4332 of $\Delta t_{AB} = 111.3 \pm 3$ is a reasonable and comprehensive value, which can be applied in lens modelling.

\section{HS~2209+1914}

\subsection{Deconvolution results and light curves}

The doubly lensed quasar HS~2209+1914 was observed by four telescopes involved in COSMOGRAIL: the Euler telescope started observations in May 2004 and was joined by the Mercator telescope, which continued the monitoring until December 2008. In May 2007, observations of this lens also started at the Maidanak Observatory during two seasons, and they were joined by the HCT the same year. This last telescope has continued observations until now. Even without the end of the 2012 season, we dispose of a total of 8.5 seasons of observations from 2004 to mid-2012.

The lensed quasar can be observed from May to December suffering from a non-visibility window of four months. The temporal sampling varies with the seasons and telescopes: combining data from Mercator, HCT, and Maidanak yields the best sampling of more than one point every second night in 2007 and 2008. This drops to one point every fortnight on average in 2010, 2011, and 2012 when we only dispose of HCT data. Table \ref{tab:HS2209_monitoring} gives an overview of the optical monitoring for this lens. One epoch corresponds to one data point in the light curve, which is the median value for that night, as explained in Section \ref{sec:photo}.

\begin{table*}[htdp]
\caption{Overview of the optical monitoring for HS~2209+1914.}
\begin{center}
\begin{tabular}{lllcclr}
\hline
\hline
Telescope & Location & Instrument & Pixel scale & Field & Time span  & Epochs  \\
\hline
Euler 1.2-m & ESO La Silla, Chile & C2 &  $0\farcs344$ & $11\arcmin \times 11\arcmin$ & May 2004 - Oct 2006 & 21 \\
Mercator 1.2-m & La Palma, Canary Islands, Spain  & MEROPE & $0\farcs193$ & $6\farcm5 \times 6\farcm5$ & Aug 2004 - Dec 2008 & 191 \\
AZT-22 1.5-m & Maidanak Observatory, Uzbekistan  & SITE & $0\farcs266$ & $3\farcm5 \times 8\farcm9$ & May 2008 - Aug 2008 & 32 \\
AZT-22 1.5-m & Maidanak Observatory, Uzbekistan  & SI & $0\farcs266$ & $18\farcm1 \times 18\farcm1$ & May - Oct 2007  & 58 \\
& & & & & + Sept - Oct 2008 &  \\
HCT 2m & Hanle, India & HFOSC &  $0\farcs296$ & $10\arcmin \times 10\arcmin$ & Sept 2007 -Aug 2012 & 136 \\
\hline
\end{tabular}
\end{center}
\label{tab:HS2209_monitoring}
\end{table*}

\begin{figure*}[htbp]
\begin{center}
\includegraphics[width=0.95\linewidth]{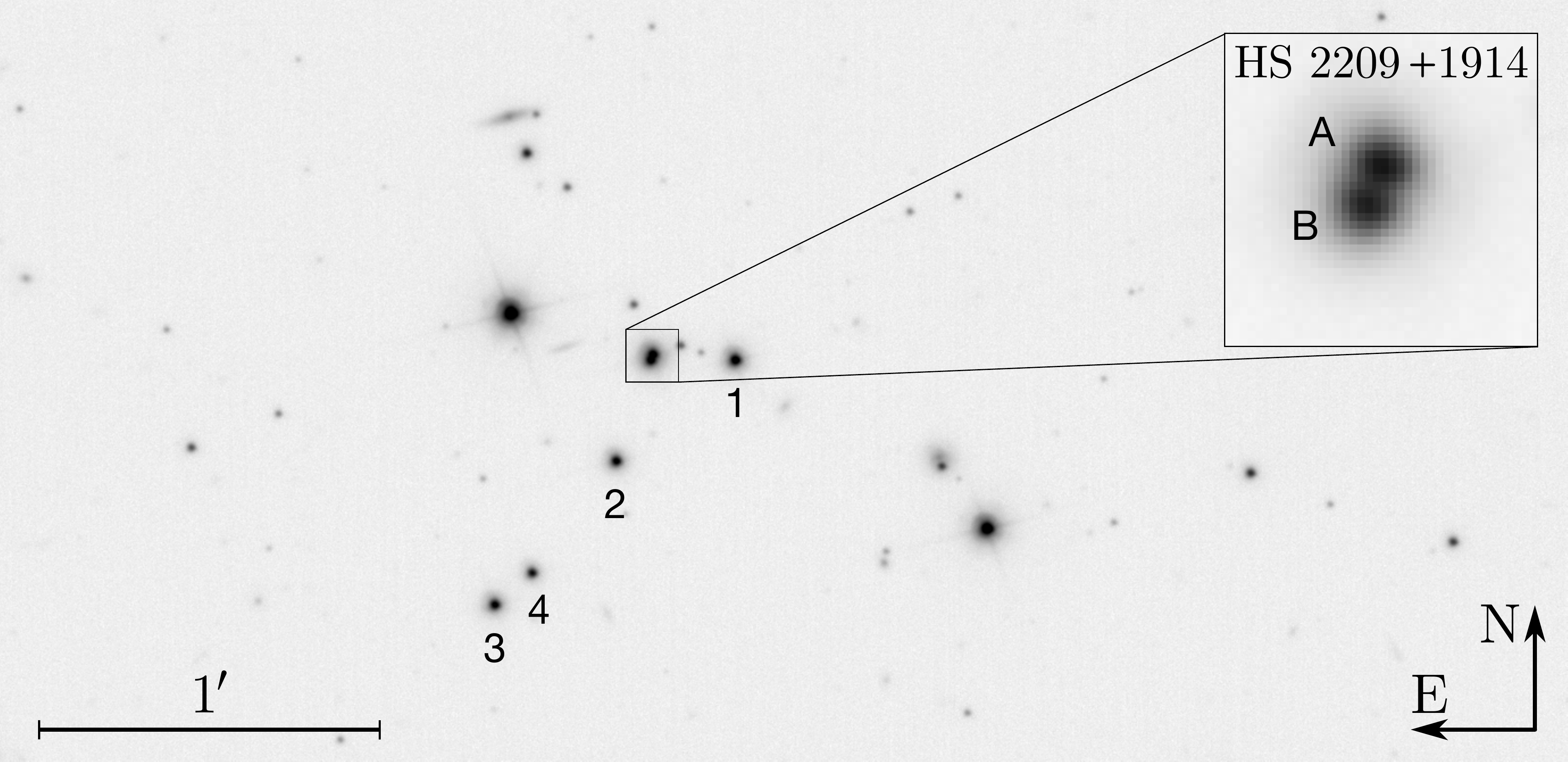}
\caption{Part of the field of view around HS~2209+1914. It is a combination of 49 frames taken at the Mercator telescope with a seeing $\leq 1\farcs2$. The stars used to model the PSF are labelled 1 to 4. All but the last one are also used to calculate the normalization coefficient. North is up and East is left.
}
\label{fig:HS2209_ref}
\end{center}
\end{figure*}

Figure \ref{fig:HS2209_ref} shows the four stars, labelled 1 to 4, used to model the PSF for the simultaneous deconvolution. Three of these stars, 1 to 3, were also used as flux calibrators.

\begin{figure}[htbp]
\begin{center}
\includegraphics[width=0.9\linewidth]{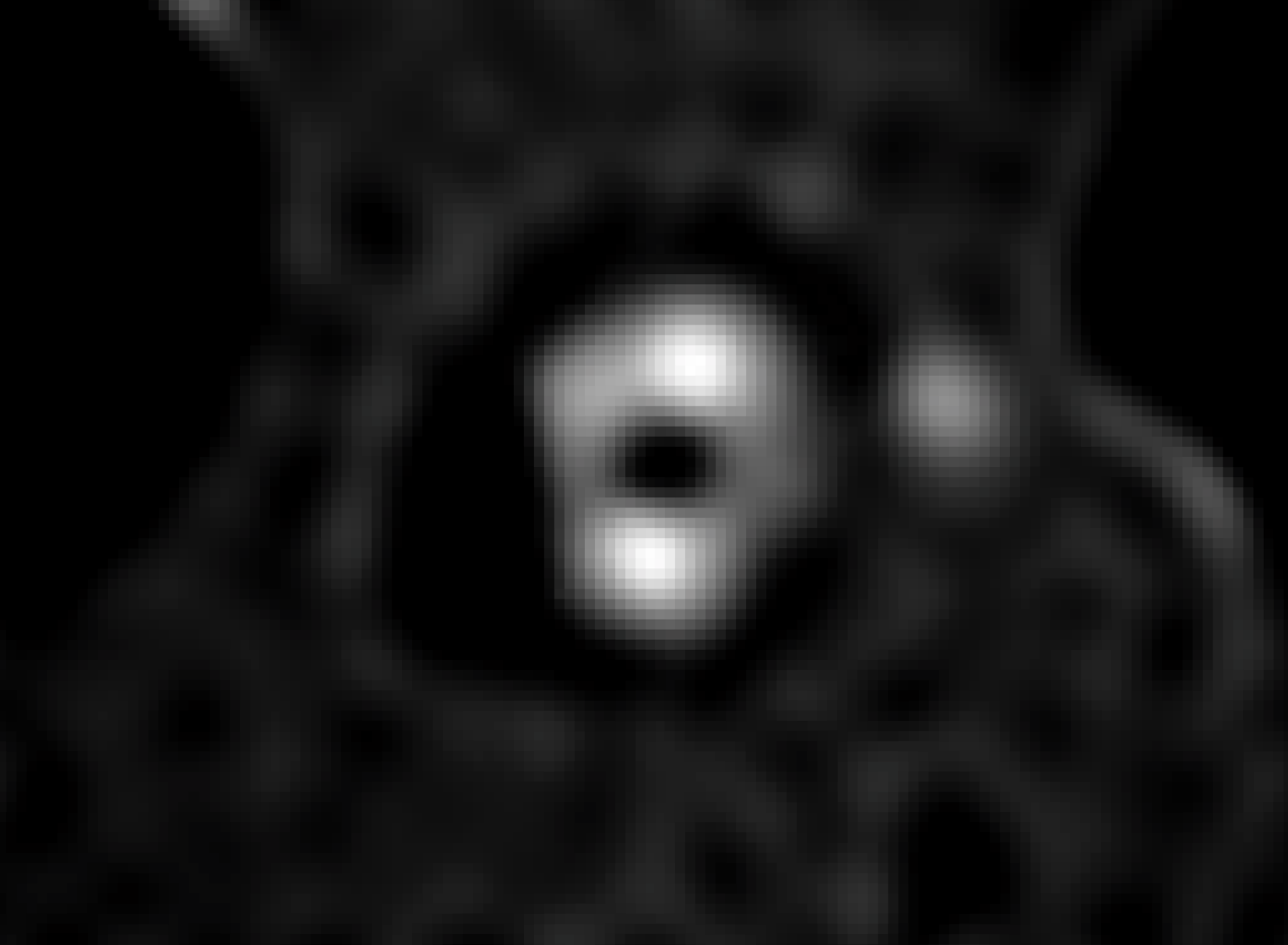}
\caption{The numerical background common to the simultaneous deconvolution of 2242 R-band images of HS~2209+1914 reveals a faint ring-like structure that is very similar to the deconvolved Hubble Space Telescope images shown in \cite{Chantry2010}. A secondary galaxy to the west of the system is also visible. However, the main lensing galaxy cannot be resolved given the compactness of the system. North is up and east is left.}
\label{fig:HS2209_back}
\end{center}
\end{figure}

In order not to degrade our photometry by some bad exposures, we eliminated all images with a seeing $>3\arcsec$ or an ellipticity of the point sources $>0.3$, as well as images with too high a sky level or deviating normalization coefficient. Simultaneous deconvolution of all remaining 2242 images reveals a faint ring-like structure (see Figure \ref{fig:HS2209_back}) that is very similar to the deconvolved Hubble Space Telescope images shown in \cite{Chantry2010}. A secondary galaxy to the west of the system is also visible. 

However, the main lensing galaxy cannot be resolved, given the compactness of the system. Both the image separation of only $1\farcs02$ and the flux ratio after deconvolution of $(F_{B}/F_{A}) = 0.82$ agree with \cite{Chantry2010}, who find $1\farcs04$ and $0.79 \pm 0.027$, respectively, even if our flux ratio has not been corrected for the time delay yet.

\begin{figure*}[htbp]
\resizebox{\hsize}{!}{\includegraphics{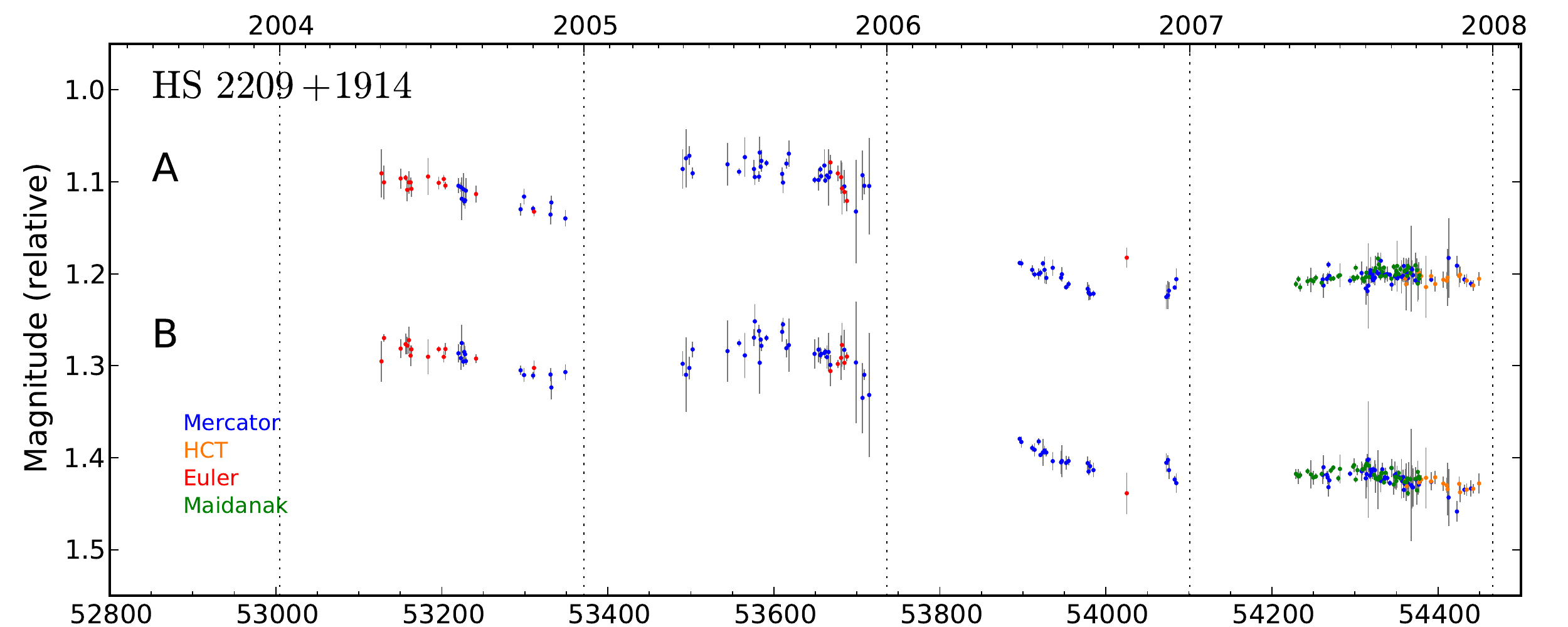}}
\resizebox{\hsize}{!}{\includegraphics{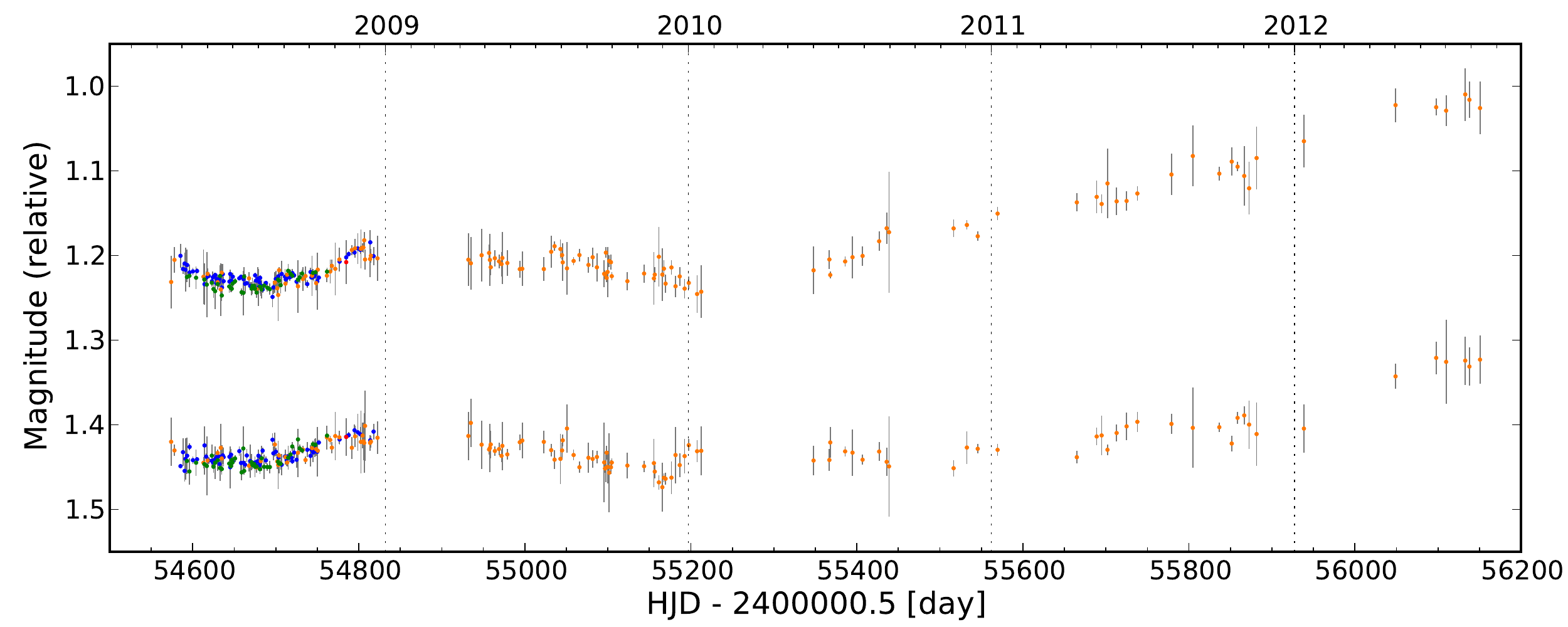}}
\caption{Light curves of HS~2209+1914 combining data from Mercator, HCT, Euler, and Maidanak and spanning nearly 9 observing seasons. Photometry is obtained from simultaneous deconvolution, and the $1\sigma$ error bars are calculated as described in Section \ref{sec:photo}.}
\label{fig_lcHS2209}
\end{figure*}

Applying the data reduction described in Section \ref{sec:photo}, we show the complete light curve in Figure \ref{fig_lcHS2209}. Euler and Mercator data complement each other well in the 2004 and 2005 seasons. The 2007 and 2008 seasons show a very good match between the data from the Mercator, Maidanak, and HCT telescopes. Even if we can observe brightness variations in the light curve up to 0.2 mag over the years, clear features within a season are unfortunately absent, which complicates the time-delay measurement. However, the slight rise in magnitude at the end of 2009 in the B curve pins down the order, with the fluctuations in the B image preceding those observed in A.

\subsection{Time delay}

\begin{figure}[htbp]
\begin{center}
\includegraphics[width=0.9\linewidth]{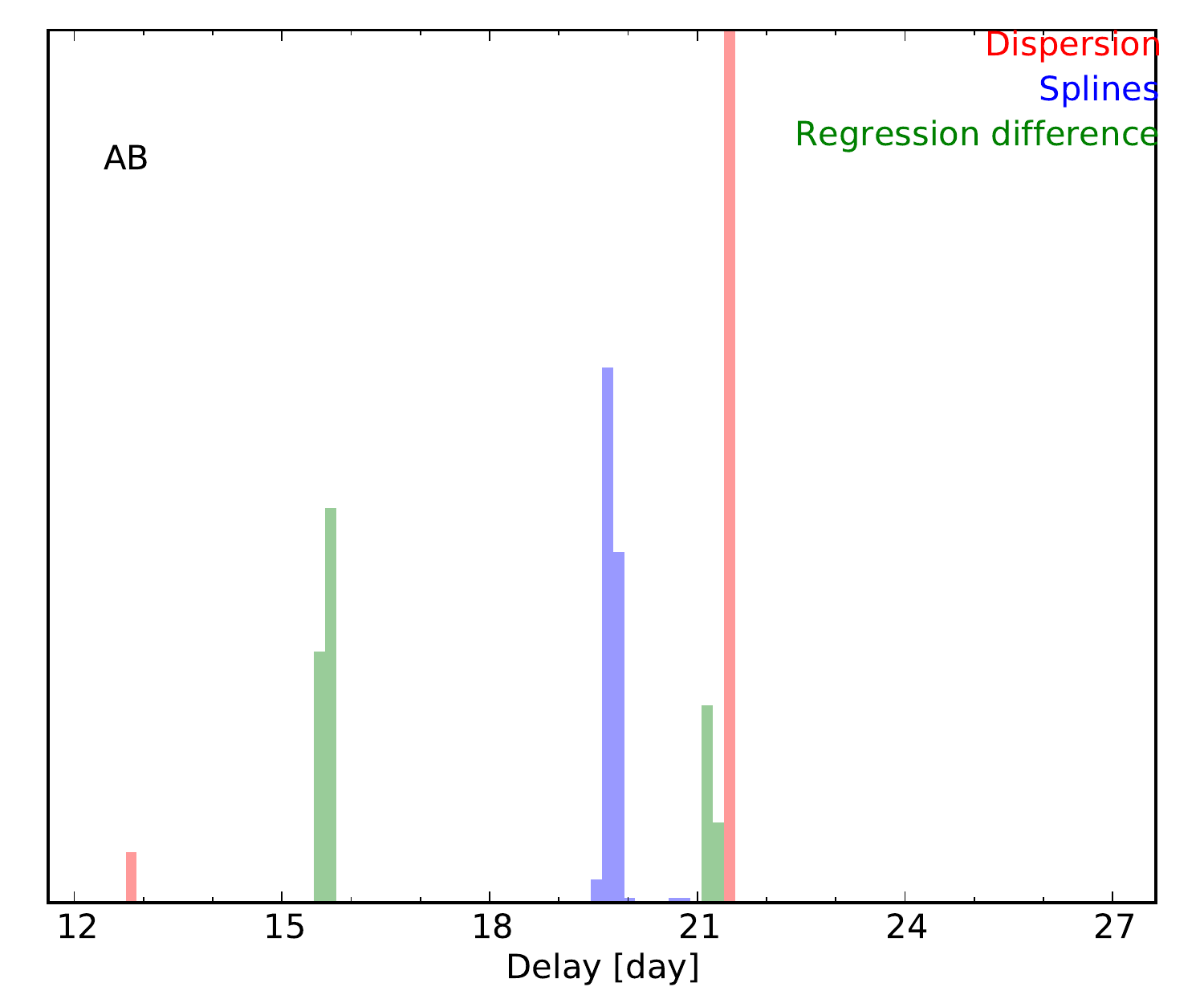}
\caption{Distribution of time delays for HS~2209+1914 obtained by running three of the time-delay estimation techniques on the complete data set 200 times, while uniformly and randomly varying the initial delay in a range of $\pm10$ days around our first guess for the time delay.}
\label{fig:HS2209_intrinsvar}
\end{center}
\end{figure}

No theoretical estimations or previous measurements of the time delay are available for this system. Our analysis is based on the light curves shown in Figure \ref{fig_lcHS2209}, and it makes use of the four techniques of which we recalled the basic principles in Section \ref{sec:photo}. As we explained in more detail in the previous section on SDSS~J1206+4332, we first test the sensivity of the results to the initial guess of a time-delay value necessary for the dispersion, regression, and spline fit methods. In Figure \ref{fig:HS2209_intrinsvar} we see that the regression difference technique and, to a lesser extent, the dispersion method show a second solution. We take this into account because our final best time delay is the weighted mean of these different values.

\begin{figure}[htbp]
\begin{center}
\includegraphics[width=0.9\linewidth]{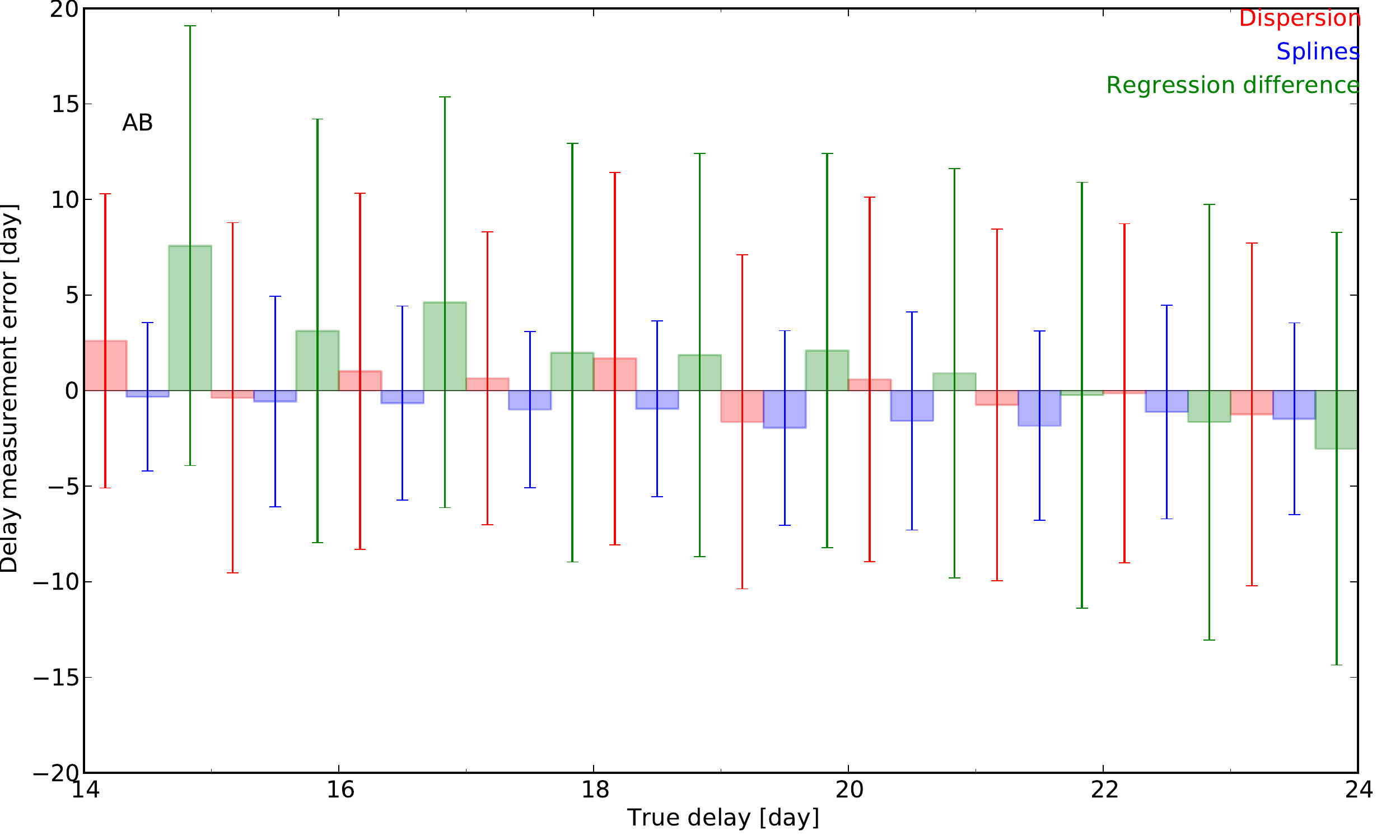}
\caption{Error analysis of the simulated light curves for HS~2209+1914 using all data from Euler, Mercator, Maidanak and HCT. We plot binned mean measurement errors and their standard deviations against the true delays of the synthetic curves. The random errors largely dominate the systematic ones, and are the smallest for the spline fit method.}
\label{fig:HS2209_measvstrue}
\end{center}
\end{figure}

The error bars for the dispersion, regression difference, and spline fit techniques are calculated in the same way as in the previous section. The analysis of the random and systematic errors, based on 1000 synthetic curves with known time delays, and presented in Figure \ref{fig:HS2209_measvstrue}, show very small systematic errors in comparison to the large random errors. 

Applying the numerical model fit to our light curve constitutes an independent test for the time-delay results. To have realistic error bars on our time delay, we have to check, as previously explained, whether the residuals from our fit are compatible with a normal Gaussian distribution with appropriate mean and sigma. This is not the case for HS~2209+1914, so we multiply the photometric error bars of the data by the appropriate positive factor to obtain a correct distribution. As mentioned in \citet{Tewes2012TD} (submitted), a rescaling of these error bars does not directly influence the uncertainties on the time-delay measurement for the dispersion, regression difference, and spline fit techniques, but it does of course for the NMF method as the noise added to the model light curve directly depends on the photometric error bars of the data points. Larger photometric errors lead to a larger uncertainty on the time delay obtained by the NMF.  

According to the telescopes involved, the NMF method finds a flux ratio corrected by the time delay ranging between $F_{B}/F_{A} = 0.82 $ (Maidanak) and $F_{B}/F_{A} = 0.85 $ (Mercator) and of $F_{B}/F_{A} = 0.84 $ when applied to the complete data set. These values are only slightly higher than the ratio obtained after deconvolution mentioned previously.

We present an overview of our time-delay results from the four different methods on the complete data set in Table \ref{tab:TD_HS2209}. Image B leads image A by $\Delta t_{BA}$ days. We notice that in spite of very large error bars, especially for the dispersion and regression difference techniques, the different time-delay estimators do converge to a common time delay of $\Delta t_{BA} \approx 20$ days with image B leading A.

\begin{table}[htbp]
\begin{center}
\begin{tabular}{ll}
\hline
\hline
Method & $\Delta t_{BA}$ in days \\ 
\hline

Dispersion & $20.99 \pm 10.09$ (9.74, 2.60)  \\
Regression  & $20.08 \pm 13.78$ (11.50, 7.58)  \\
Spline & $19.77 \pm 6.03$ (5.71, 1.95)  \\
NMF & $19.28 \pm 1.48$ \\

\hline
\end{tabular}
\end{center}
\caption{Summary of time delays for HS~2209+1914 obtained by the four different methods on the complete light curve with data from Euler, Mercator, Maidanak, and HCT. The indicated error is the total error. Between brackets we mention the random and the systematic error for the first three methods as explained in the text.}
\label{tab:TD_HS2209}
\end{table}

The large error bars for two of our methods can be partially explained if we take a closer look at the contributions of the data from different telescopes to the time delay. We applied the NMF to five subsets of data: 1) Mercator only, 2) Maidanak only, 3) HCT only, 4) Mercator + HCT, 5) Euler + Mercator + HCT. Comparing the results of a single run of the method on these subsets and a Monte Carlo approach of 1000 runs on perturbed model curves, both presented in Table \ref{tab:TD_NMF_HS2209}, where image B leads image A by $\Delta t_{BA}$ days, reveals interesting information. Mercator and HCT data do not converge to the same time delay: Mercator data point to one around $\approx 23$ days, whereas HCT data allow a possible delay around $\approx 33$ days. With Maidanak data, which cover the two overlapping seasons between Mercator and HCT exactly, we find both delays. With bare Maidanak data, the lower Mercator value is preferred, but once we perturb the model curve, we find a bimodal distribution with peaks around $\approx 22$ and $\approx 34$ days. The more data we add to the light curve, the more the lower delay dominates. 

\begin{table*}[htbp]
\begin{center}
\begin{tabular}{lcc}
\hline
\hline
Subset & $\Delta t_{BA}$ in days & $\Delta t_{BA} \pm$ err \\ 
& (single run) & (Monte Carlo approach) \\
\hline

Mercator & 23.1 &$23.20 \pm 2.96$  \\
Maidanak & 21.9 &$30.48 \pm 9.32$    \\
HCT & 31.8 & $33.32 \pm 4.04$    \\
Mercator + HCT & 25.9 &$26.50 \pm 1.95$  \\
Euler + Mercator + HCT & 22.7 & $26.18 \pm 1.86$ \\
Euler + Mercator + Maidanak + HCT & 19.6 & $19.28 \pm 1.48$ \\

\hline
\end{tabular}
\end{center}
\caption{Summary of time delays for HS~2209+1914 obtained by the NMF on five subsets of data and on the complete set.}
\label{tab:TD_NMF_HS2209}
\end{table*}

The remarks we made at the end of Section \ref{subsec:td_J1206} on the comparison of the results are also valid for the time delay in HS~2209+1914. However, the presence of a possible second solution in the data and the question of whether the error bars should reflect this presence complicate the picture.
Again, basically different methods converge towards the same value of $\Delta t_{BA} \approx$ 20 days, so a final result of $\Delta t_{BA} = 20 \pm 5$ is a reasonable and intermediate solution between possibly too optimistic and too pessimistic error bars, while not completely excluding higher values since these uncertainties are $1\sigma$ error bars. However, given the lack of obvious fast intrinsic quasar variability features in these light curves, and the possible presence of microlensing, this time-delay measurement should be handled with caution.



\section{Conclusions}

In this paper we present seven-season-long light curves of two doubly lensed quasars based on COSMOGRAIL observations combining data from Mercator, Maidanak, and HCT telescopes for SDSS~J1206+4332, as well as with data from Euler for HS~2209+1914. All data were reduced in a homogeneous way described in \citet{Tewes2012_RXJ1131} (submitted). We applied four time-delay estimation methods, based on different principles, on these light curves. 

For SDSS~J1206+4332 simultaneous deconvolution reveals the three galaxies that possibly contribute to the lensing potential of the system. We consider the results of the spline fit method as our final best time delay of 111 days and pin down the error bars to 3.5\%, thus improving the uncertainty over a previously published result \citep{Paraficz2009}. 

Hardly anything has been published on the lensed quasar HS~2209+1914 since its discovery. Our work confirms the image separation, the flux ratio, and a faint ring-like structure as mentioned in \citet{Chantry2010}. We are the first to measure a time delay of $\Delta t_{BA} \approx 20$ days, which was only possible by combining data from four telescopes into a 8.5-year-long light curve.

Even if quadruply lensed quasar systems allow a more straightforward estimation of the Hubble Constant $H_{0}$, \cite{Suyu2012a} has shown that doubly lensed systems can augment the number of useful time delays when high resolution images of the source galaxy in extended arcs can break the radial profile slope degeneracy. HS~2209+1914 is promising in this respect even if the compactness of the system and the short time delay make it difficult to reach sufficiently high precision.

\begin{acknowledgements}
We are deeply grateful to the numerous observers who contributed to the data acquisition at the Flemish-Belgian 1.2-m Mercator telescope (La Palma, Spain), the Swiss 1.2-m Euler telescope (La Silla, Chile), the Uzbek 1.5-m AZT-22 telescope (Maidanak, Uzbekistan), and the Indian 2.0-m HCT telescope (Hanle, India). 

COSMOGRAIL is financially supported by the Swiss National Science Foundation (SNSF). 

This work was partially supported by ESA and the Belgian Federal Science Policy (BELSPO) in the framework of the PRODEX Experiment Arrangement C-90312. 


\end{acknowledgements}

\bibliographystyle{aa} 
\bibliography{PhDbiblio}
\end{document}